# Towards understanding the immune system

E.Ahmed* and A.H. Hashish**
*Department of Mathematics, Faculty of Science, Mansoura University,
Mansoura, EGYPT.
**Department of Physics, Faculty of Science, UAE University,
P.O.Box 17551 Al-Ain, UAE

**Abstract**: It is proposed that using both self-non-self and danger theories give a better understanding of how the immune system works. It is proposed that comparing immune system to police force is useful in this case since police responds both to danger or damage signals and to foreign or suspicious behavior even if no danger signals existed. We also propose that due to low zone tolerance immunotherapy needs to be combined with another treatment method for cancer e.g. chemotherapy or/and radiotherapy to get a sufficient eradication of tumors. Finally we propose that fractional order differential equations are more suitable than the familiar integer order differential equations. A fractional order example of two immune effectors attacking an antigen is given
.

## 1. Introduction:

Biology is a rich source for mathematical ideas. Also mathematics can help in understanding some aspects of the biological phenomenon. Here we study two aspects of the immune system (IS) [Segel and Cohen, 2001]. The first is the initiation of IS and the second is how to model IS. So far two theories are competing to explain the initiation of IS. The the first is self-nonself theory [Janeway, 1992] and the second is th danger theory [Matzinger, 1998]. In sec.2 we propose that they complement each other and that both of them should be used. In sec.3 we argue that fractional order differential equations [Stanislavsky, 2000] are more suitable to model IS than integer order ones. We give an example of two immune effectors attacking an antigen (Ag) e.g. a virus or a microbe.

## 2. Initiation of immune system

Immune system (IS) is one of the most fascinating schemes from the point of view of biology, physics, computer science and mathematics. The immune system is complex, intricate and interesting (Ahmed and Hashish, 2005). It is known to be multi-functional and multi-pathway i.e. most immune effectors do more than one job. Also each function of the immune system is typically done by more than one effector. This makes it more robust.

The immune system has a series of dual natures, the most important of which is self-non-self recognition. The others are: general/specific, natural/ adaptive, innate/ acquired, cell-mediated/humoral, active/passive, primary/secondary. Parts of the immune system are **antigen-specific** (they recognize and act against particular antigens), **systemic** (not confined to the initial infection site, but work throughout the body), and have **memory** (recognize and mount an even stronger attack to the same antigen the next time, Ahmed and Hashish, 2003). Sometimes the process breaks down and the immune system attack self-cells. This is the case of **autoimmune diseases** like multiple sclerosis.

Understanding the initiation and maintenances of IS is an interesting problem. For IS, self-non-self recognition is achieved by having every cell display a marker based on the major histocompatibility complex (MHC). Any cell not displaying this marker is treated as non-self and attacked. Originally it was self-non-self theory (SNS) by Janeway (1992) which states that it is the foreignness of the antigens which stimulate the IS to attack them.

Despite its successes this theory has several problems: Firstly, self is variable with time. Secondly, it was thought that self reactive cells are removed from the thymus (a process called negative selection). But it was discovered that removing thymus from adults does not significantly increase autoimmune diseases (IS attacking self). Thus negative selection may be more effective for young children than for adults. Other mechanisms e.g. high and low zone tolerance i.e. immune effectors do not attack Ag whose number is too high or too low. Moreover; the two signal system helps in controlling autoimmune diseases since immune effectors have to receive two signals one indicating the existence of Ag and the other stimulating the effector. The absence of the second signal. Thirdly, why IS tolerates harmless foreign Ag? After all they are non-self.

Recently, Matzinger [1998 and 2002] proposed an interesting theory namely danger theory for the initiation of IS [Matzinger 1998, 2002]. According to this theory IS reacts if it receives danger signals no matter what caused it, hence self-non-self discrimination is not required. Again despite several successes and its elegance one thinks that this theory has some problems: One can't consider MHC I, II as a kind of self markers?. Also if only the danger signals initiate IS then why some organs are rejected after a long time of being transplanted i.e. after the danger signals should have disappeared? Moreover, danger signals in this model do not include starved cells or cells under pressure e.g. within or near a tumor. In fact it assumes that IS does not react to tumors at all. Finally, in some cases activated immune effectors do not require co-stimulation (second signal) from antigen presenting cells (APC) which in the danger theory are the ones that deliver danger signals.

We propose that the two models compliment each other. We argue for this in several ways:
   **The first** is that, as stated at the beginning, IS is a multi-pathways system hence its initiation is expected to be by several mechanisms not just one.

   **The second** is that we think a useful analogy of IS is to police in a city. It is known that police responds to danger signals but at the same time it responds to suspicious behavior even if there are no danger signals.

   **The third** is that by combining both theories one is able to explain several phenomena which may cause problems if one uses only one theory.

Now we attempt to see how the combined (Police Theory) is capable of explaining several phenomena of the immune system:

I) **Transplantation**: If the transplanted organ carried non-self marker then it should be rejected. If the danger signals reach IS then again it will be rejected. After a long time of transplantation the first mechanism still works hence the negative response of long-time transplantation rejection is explained.
The relative success of liver transplantation may be explained by high zone tolerance (high and low zone tolerance is the phenomenon that the immune system reacts only if the antigen number is neither too large nor too small) since this organ is large and quick to regenerate.

II) **Tumors:** They by definition do not respond to control signals. Thus their receptors have to be deformed somehow. This may make them detectable as non-self. Therefore, we expect that immune response to some tumors exist. Due to the low zone tolerance we expect that immunotherapy alone will not sufficiently eradicate a tumor. We propose that it should be combined with another treatment e.g. chemotherapy and/or radiotherapy.

III) **Auto-reactivity:** It is expected to occur due to the randomness of the immune receptors (or autoimmunity occurs when the immune system attacks the body it is supposed to protect). Hence several mechanisms exist to control it. The first is the negative selection in the thymus. However, it is known that adults whose thymuses are removed do not show a significant increase in autoimmunity. Thus other effective control mechanisms exist, e.g. high and low zone tolerance which prevents self reactive effectors from attacking large or very small organs. Also the two signal mechanism is a quite effective one against autoimmunity since immune effectors react only if it receives two signals; the first indicating danger or non-self and the second, co-stimulates it. Typically antigen presenting cells do not deliver the second signal to self reactive cells. At the same time the existence of useful natural self reactive cells [Matzinger, 1998] do not contradict either self-non-self or danger theories.

## 3. Fractional order equations (FE) and the immune system:

**Definition (1)**[Smith 2003]: A complex adaptive system (CAS) consists of inhomogeneous, interacting adaptive agents.
**Definition (2):** "An emergent property of a CAS is a property of the system as a whole which does not exist at the individual elements (agents) level". Typical examples are the brain, the immune system, the economy, social systems, ecology, insects swarm, etc…

Therefore to understand a complex system one has to study the system as a whole and not to decompose it into its constituents. This totalistic approach is against the standard reductionist one, which tries to decompose any system to its constituents and hopes that by understanding the elements one can understand the whole system. Recently [Segel and Cohen 2001, Ahmed and Hashish 2005] this idea has been called for in the immune system.

Recently [Stanislavsky, 2000] it became apparent that fractional equations naturally represent systems with memory. To see this consider the following evolution equation:

$$df(t)/dt = -\lambda^2 \int_0^t k(t-t')f(t')dt' \qquad (1)$$

If the system has no memory then $k(t-t') = \delta(t-t')$ and one gets $f(t) = f_0 \exp(-\lambda^2 t)$. If the system has an ideal memory then $k(t-t') = \{1 \text{ if } t \geq t', 0 \text{ if } t < t'\}$ hence $f \approx f_0 \cos \lambda t$. Using Laplace transform $L[f] = \int_0^\infty f(t)\exp(-st)dt$ one gets L[f]=1 if there is no memory and 1/s if there is ideal memory hence the case of non-ideal memory is expected to be given by $L[f] = 1/s^\alpha$, $0 < \alpha < 1$. In this case equation (1) becomes

$$df(t)/dt = \int_0^t (t-t')^{\alpha-1} f(t')dt'/\Gamma(\alpha) \qquad (2)$$

where $\Gamma(\alpha)$ is the Gamma function. This system has the following solution:
$f(t) = f_0 E_{\alpha+1}(-\lambda^2 t^{\alpha+1})$,
where $E_\alpha(z)$ is the Mittag-Leffler function given by

$$E_\alpha(z) = \sum_{k=0}^\infty z^k / \Gamma(\alpha k + 1)$$

It is direct to see that $E_1(z) = \exp(z), E_2(z) = \cos z$.

Following a similar procedure to study a random process with memory, one obtains the following fractional evolution equation

$$\partial^{\alpha+1} P(x,t)/\partial t^{\alpha+1} = \sum_n (-1)^n \partial^n [K_n(x)P(x,t)]/\partial x^n / n!, \qquad 0 < \alpha < 1 \quad (3)$$

where P(x,t) is a measure of the probability to find a particle at time t at position x.
For the case of fractional diffusion equation the results are
$\partial^{\alpha+1} P(x,t)/\partial t^{\alpha+1} = D\partial^2 P(x,t)/\partial x^2$, $P(x,0) = \delta(x), \partial P(x,0)/\partial t = 0 \Rightarrow$

$$P = (1/(2\sqrt{D t^\beta}))M(|x|/\sqrt{D t^\beta}; \beta), \quad \beta = (\alpha+1)/2 \qquad (4)$$

$$M(z; \beta) = \sum_{n=0}^\infty [(-1)^n z^n /\{n!\Gamma(-\beta n + 1 - \beta)\}]$$

For the case of no memory $\alpha = 0 \Rightarrow M(z; 1/2) = \exp(-z^2/4)$.

Moreover it has been proved that fractional order systems are relevant to fractal systems and systems with power law correlations [Rocco and West, 1999]. Thus fractional equations naturally represent systems with memory and fractal systems consequently they are relevant to the immune system.

An example to show the suitability of fractional order systems to model IS consider two immune effectors **y,z** attacking an antigen **x**. We used two effectors since IS is a multi-pathways i.e. almost every function is done by more than one effector. Thus one has:

$$D^\alpha x = x - axy - bxz, \quad D^\alpha y = -cy + xy, \quad D^\alpha z = -ez + xz \quad \quad (5)$$

where a,b,c,e are positive constants and $\alpha \in [0,1)$. There are four equilibrium solutions for **(5)**: I) (0,0,0), II) (c,1/a,0), III) (e,0,1/b) and IV) the internal solution where (x,y,z) are nonzero. But this implies **e = c**. Since we assume that the effectors y,z are different then **e<>c** hence solution IV) is not allowed.

State I is the naïve state. State II, III are memory states where both the antigen and an effector persists provided that in II (III) the parameter c (e) is less than or equal to the immune system threshold. This represents the major mechanism of memory namely the antigen persistent mechanism. This mechanism depends on the low zone tolerance property of the immune system.

The stability of fractional order differential equations has been pioneered by Matignon [Matignon 1996] and where he derived that an equilibrium of a linear system is locally asymptotically stable if

$$|\arg(\lambda)| > \alpha\pi/2 \quad \quad (6)$$

where $\lambda$ is the eigenvalue of the Jacobian matrix of the system at the equilibrium. Applying this result to **(5)** one gets (0,0,0) is unstable. The equilibrium (c,1/a,0) is locally asymptotically stable if **e>c** since the arguments of the three eigenvalues are $\pi/2, -\pi/2, -\pi$. Similarly the equilibrium (e,0,1/b) is locally asymptotically stable if e<c. Accordingly one always gets a locally asymptotically stable equilibrium state which corresponds to the antigen persisting (x>0) memory state of the immune system. When one uses integer order equations there is no locally asymptotically states for **(5)**.

## 4. Conclusions:

In this paper we concluded the following:
1) Both self-non-self and danger theories should be used to describe the initiation of the immune system.
2) Immunotherapy should be augmented by other modes of treatments for cancer since typically IS cannot completely eradicate the tumor.
3) Fractional order dynamical systems are more suitable to model the immune system than their integer order counterpart.

**Acknowledgmen**t: We thank the referee for comments.

## References:

Ahmed E. and A H Hashish (2005), **On modelling the immune system as a complex system**, Theory in Biosciences, **124** (in press )

Ahmed E.and A H Hashish (2003), **On modelling of immune memory mechanisms**, Theory in Biosciences, Vol **122**(4): pp 339-342.

Janeway C.A. Jr (1992), "**The immune system discriminates infectious non-self from non infectious self**", Immunology Today, **13**, 11.

Matignon D. (1996) , **Stability results for fractional differential equations with applications to control processing**, Computational Eng. in Sys. Appl. Vol.2 Lille France 963.

Matzinger P. (1998), **An innate sense of danger,** Seminars in immunology, **10**, 399.

Matzinger P. (2002), **The danger model. A renewed sense of self**, Science, **296**, 301.

Rocco A. and West B.J. (1999), "**Fractional Calculus and the Evolution of Fractal Phenomena**"' Physica A 265, 535.

Segel L.A. and Cohen I.R.(eds.)(2001),"Design principles for the immune system and other distributed autonomous systems", Oxford Univ.Press U.K.

Smith J.B., "**A technical report of complex system**", CS 0303020, 2003

Stanislavsky A.A.(2000), "**Memory effects and macroscopic manifestation of randomness"**, Phys. Rev.E61, 4752.